# Squeezed Metallic Droplet with Tunable Kubo Gap and Charge Injection in Transition Metal Dichalcogenides

Jiaren Yuan,[1,2] Yuanping Chen,[1] Yuee Xie,[1] Xiaoyu Zhang,[1] Dewei Rao,[1] Yandong Guo,[3] Xiaohong Yan,[1*] Yuanping Feng,[2*] & Yongqing Cai[4*]

[1]College of Science and School of Material Science and Engineering, Jiangsu University, Zhenjiang, 212013, China;

[2]Department of Physics and Centre for Advanced Two-Dimensional Materials, National University of Singapore, 2 Science Drive 3, 117551, Singapore;

[3]College of Electronic Science and Engineering, Nanjing University of Posts and Telecommunications, Nanjing 210046, China;

[4]Joint Key Laboratory of the Ministry of Education, Institute of Applied Physics and Materials

Engineering, University of Macau, Taipa, Macau, China.

Correspondence author: X.Y.(email: yanxh@ujs.edu.cn); Y.F. (email: phyfyp@nus.edu.sg); Y.C.(email: yongqingcai@um.edu.mo)

Shrinking the size of a bulk metal into nanoscale leads to the discreteness of electronic energy levels, the so-called Kubo gap $\delta$. Renormalization of the electronic properties with a tunable and size-dependent $\delta$ renders fascinating quantum emission and tunneling. Breaking away from traditional trend of focusing almost exclusively on three-dimensional (3D) metal clusters, here we **demonstrate that Kubo gap δ can be achieved with a two-dimensional (2D) metallic transition metal dichalcogenide (i.e., 1T'-phase $MoTe_2$) nanocluster embedded in a semiconducting polymorph (i.e., 1H-phase $MoTe_2$). Such a 1T'/1H $MoTe_2$ nanodomain resembles a 3D metallic droplet squeezed in a 2D space which shows a strong polarization catastrophe while simultaneously maintains its bond integrity which is absent in traditional δ-gapped 3D clusters. The weak screening of the host 2D $MoTe_2$ leads to quantum emissions of such pseudo-metallic systems and a ballistic injection of carriers in the 1T'/1H/1T' homojunctions which may find applications in sensors and 2D reconfigurable devices.**

Quantum confinement and surface effect induce non-scalable and highly non-monotonic physicochemical properties of nanoparticles ranging from atom to several nanometers[1]. The electronic structure of a nanoparticle strikingly depends on the size. As the size of a metallic nanoparticle is reduced, its extending electronic wavefunction becomes quantum confined and evolves into shell-like behaviors, i.e. discretization of energy levels[2]. The average spacing of the successive levels, known as the Kubo gap δ, scales with $E_f/N$ where $E_f$ and $N$ are the Fermi energy of the bulk metal and the nuclearity of the particle, respectively[3]. For a Ag nanoparticle of 3 nm in diameter ($N \sim 10^3$ atoms), δ is 5–10 meV[4] while for a Na nanoparticle of 2.4 nm, δ would be ~26 meV[3]. Striking phenomena occur when δ is greater or comparable to thermal energy $k_BT$ (at room temperature, $k_BT$=25 meV) which renders its nonmetallicity[5]. The occurrence of the Kubo gap in metallic nanoparticles also accounts for other fascinating properties such as significantly lower melting points[6], nonmagnetic-magnetic transitions[7] and changes in spectral features[8]. However, the moderate δ gapped metallic particles tend to suffer from multiple structural variations, i.e. isomerization of Ag and Au nanoparticles of different charging states[9,10]. The multivalence of these nanoclusters leads to varying structures[11], unstable magnetic ordering and spin excitation[12] which severely hinder their applications. Because of the dramatic structural discontinuity at the particle-vacuum/liquid interface of these metallic nanoparticles it is hard to reach true monodispersity due to structural agglomerations and reconstruction.

Here we demonstrate a new scheme for the realization of a Kubo gap in a lattice-continual metallic nanophase embedded in a semiconducting host using transition metal dichalcogenides (TMDs) $MX_2$ (M: Mo or W and X: S, Se, or Te). These TMDs have a strong lattice, spin, orbital coupling and a wealth of polymorphs with semiconducting (1H), metallic (1T and 1T') phases which trigger significant interests in catalysis and nanoelectronics[13-16]. Many of these TMDs, stabilizing in a hexagonal (1H) phase, show a semiconducting characteristic with a strong spin-orbit coupling and excitonic effects, which are suitable for diverse applications in field effect transistor[17], magnetic tunnel junction[18], valleytronics[19] and optoelectronic devices[20]. Unlike the 1H phase, the metallic octahedral 1T and its distorted octahedral 1T' phase exhibit a large magnetoresistance[21], intriguing quantum spin Hall effect[22] and high catalytic activities[23,24]. Amongst the various TMDs, $MoTe_2$ is particularly interesting due to the smallest free energy difference (~40 meV per unit cell[25,26]) between the semiconducting 1H phase and metallic 1T′ phase[27]. Recent theoretical work revealed a laser-induced mechanism of vacancy ordering and growth of 1T' seed in the transition[28]. By taking advantage of this phase tunability, here we demonstrate that through creating the 1T'/1H $MoTe_2$ coplanar heterophase structure, conducting carriers confined in 1T' $MoTe_2$ nanodomain show a Kubo gap opening. Different from the traditional 3D metallic nanoparticles which have undesirable surface dangling bonds, the conducting carriers in the 2D nanodomain which mimicks a metallic droplet, are squeezed into atomically thin 2D triangular space with well-passivated edge atoms. It is potentially useful for robust creation/injection of quantum particles with a weak screening.

**Results**

**Reversible phase transition induced by charge doping and strain.** Group VI TMDs with the chemical formula $MX_2$ (M: Mo or W and X: S, Se, or Te) have a variety of polymorph structures, such as the honeycomb 1H, 1T and monoclinic 1T'-phases (Fig. 1 a-c) with space group of P6m2, P3m1 and P21/m, respectively. The optimized lattice constants and the energetics, relative to the 1H phase, are given in Supplementary Table S1. Governed by the crystal field-induced splitting of the *d* orbitals of the M cation, a facile transformation from 1H to 1T/1T' occurs, driven by increasing electron density of the M orbitals from $d^2$ to the $d^{2+x}$ via lithium exfoliation or doping. Since such transition involves a strong orbital-lattice-charge coupling with breaking octahedral symmetry of the 1H phase, below we mainly focus on transition between the high symmetric 1H and the low symmetric 1T' phase. The estimation of the 1H-1T' transition can serve as an upper limit for the metastable 1T phase in those 1H-1T and 1T-1T' process under external stress fields and chemical doping.

As shown in Fig. 1d, we explore coupled effects of the strain and doping on the thermodynamics of the phase transition of TMDs. In our work, to be consistent with the varied phases, an orthorhombic cell of $MX_2$ phases is used. Uniaxial strain $\varepsilon$ is applied along the armchair direction and the lattice constant along zigzag direction is relaxed for each strain. The doping charge density $q$ is calculated by adding/depleting electrons in the orthorhombic cell (dashed rectangles in Fig. 1a-c) and measured in the unit of $q_0$, calculated as one electron injected into the equilibrium orthorhombic cell containing two chemical formulas (for $MoTe_2$, $q_0$ is

~$0.92\times10^{15}$ e cm$^{-2}$). A positive (negative) value of $q$ represents electron (hole) doping. The phase diagram in the ($\varepsilon$, $q$) space is plotted through comparing the energies $E$ ($\varepsilon$, $q$) of the 1H and 1T' phases, which are evaluated by DFT calculation, for each of the six group VI TMDs.

Interestingly, all the strain-doping curves corresponding to the phase transition are monotonically U-shaped, indicating a synergistic effect of the strain and carrier doping. The critical value of charge doping (strain) for driving the phase transition decreases when a uniaxial strain (charge doping) is applied simultaneously. For instance, 0.13 $q_0$ of electron doping or -0.2 $q_0$ of hole doping is required for the realization of the 1H-1T' transition in MoTe$_2$ without strain. With 3% strain, for both polarities only ~0.05 $q_0$ is required to activate the transition. Our findings explain the observed long range phase transition of MoTe$_2$ due to contact with the 2D electride [Ca$_2$N]$^+\cdot$e$^-$ substrate[29]. Concerning the pure strain effect without charge doping, our calculations reveal that only ~4% strain is required to induce the 1H-1T' phase transition of the MoTe$_2$, which is consistent with the result of previous experiments[30].

For X = S and Se series, without charge doping, the critical strain for realizing the phase change is extremely high, ranging around 17%-22%, which exceeds their strength limits. Charge doping reduces the critical strain, but it (>8%) is still beyond what can be reached in real experiments. Nevertheless, our work shows that charge doping, especially electron doping, is an effective mean of reducing the relative energies of the 1H and 1T' phases. On the other hand, tensile deformation of the lattice reduces hybridization of the electronic states and shallowing the gap between the $d_{z^2}/d_{x^2-y^2,xy}$ splitting associated with a weakened ligand field effect of the chalcogens. This is the underlying root of reduction in energy difference among the phases.

Surprisingly, unlike the aforementioned S and Se-based dichalcogenides, $WTe_2$ shows an opposite phase hierarchy with the 1T' phase being more stable than the 1H phase under zero strain and without doping. $WTe_2$ can be transformed from 1T' to 1H by a compression strain of about -6%. However, in real circumstances, compression tends to induce structural rippling. In addition, according to our prediction, unintentional charge doping due to defects and adsorbates stabilizes the 1T' phase and increases the cost of phase transition. Therefore, $WTe_2$ tends to be stable and suffers less from phase variations compared with its TMD cousins. This should be the underlying reason for the $WTe_2$ being as phase stabilizers for mixing with other 1H materials[31].

The curves shown in Fig. 1d are derived based on total energies from DFT calculation at 0 K which does not include contribution of vibrational free energy. For $MoTe_2$ the energy difference between the 1H and 1T' phases (about 42 meV) is comparable with $k_B T$ which is of the order of 10 meV at finite temperature ($T$). For a more accurate prediction of the transition at finite $T$ and to obtain a complete picture of strain-charge-temperature relationship, we calculate the thermodynamic potential $U(\varepsilon, q, T)$, defined as $U(\varepsilon, q, T) = E(\varepsilon, q) + F(\varepsilon, q, T)$ where $F(\varepsilon, q, T)$ is the vibrational free energy correction under quasi-harmonic approximation[32]. The phase transition point re-evaluated using this potential at 300K indicates that the phase transition condition is relaxed by the Helmholtz free energy correction in both cases of strain and charge doping, as shown in Fig. 1d. The improvement with free energy correction has no change in the shape of the phase diagram but slightly softens the critical strains for the 1H-1T' transition.

Thus far we report the thermodynamics of the 1H-1T' energy offset of TMDs. For an appropriate prediction of the effect of doping on the kinetics process of the transition, realistic descriptors like thermal activation barrier for the transition should be obtained. According to our thermodynamics screening of the various TMDs shown above, the $MoTe_2$ shows the best facile transition under finite strain and moderate doping, thus we will only focus on activation analysis of $MoTe_2$ below. Two different pathways (1H-1T-1T' and 1H-1T'), depending on whether the intermediate 1T phase is included, are identified. By using climbing-image nudged elastic band (CI-NEB) method, the energy barriers are evaluated. The energy barrier of displacive phase transition of 1H-1T-1T' occurs by the migration of Te atoms while the Mo atoms remain fixed, and the barrier is 1.09 eV per chemical formula. In contrast, the 1H-1T' pathway involves sliding of Te and Mo atoms simultaneously and the barrier is 0.87 eV per chemical formula. Therefore, the 1H-1T' transition is more likely to take place via the cooperative process which involves displacements of both cationic and anionic atoms. This process can be affected by strain and charge doping. Further calculations reveal that a tensile uniaxial strain continuously reduces the barrier, down to 0.63 eV at 10% strain. Concerning the charge doping, the energy barrier reduces quickly for both electron and hole doping, reaching 0.52 eV for -0.5 $q_0$ hole doping and 0.60 eV for 0.5 $q_0$ electron doping. It is expected that the coupling of strain and charge doping would lead to an even lower value.

**Energetics and Kubo gap of embedded metallic quantum dot.** The deformation of the 1H/1T' lattice in the strain/doping induced phase transition discussed above is driven by the excess carriers occupying the *d*-orbital of the cations. The highly localized nature of the *d*-orbitals suggests a realization of locally controlled atomic displacement in TMDs through doping via local gating together with a proper strain engineering (i.e. via nano-indentation, see schematics in Fig. 2a). While phase engineering of $d^2$-type TMDs has been previously demonstrated[26, 33-35], realization of nano-patterning of these semiconducting 1H- metallic 1T' phase in a single sheet is still challenging owing to the inherent complex boundaries[36]. Here we show that formation of confined nano 1T' $MX_2$ phase, resembling normal 3D nanocluster/dots but squeezed in 2D, creates a sizable and tunable Kubo gap. Owing to the small energy difference between the 1H and 1T' phase of $MoTe_2$ which benefits formation of stable heterostructures, we focus on $MoTe_2$ in the following discussion.

Triangular shaped metallic 1T' phase domain of $MoTe_2$ is modelled since the 60° intersected boundaries dominate in the 1H/1T/1T' $MX_2$ sheet[37]. In contrast to traditional metallic clusters which can show symmetries (i.e. fivefold symmetric operations) different from their bulk counterparts, here the 1H/1T' domains have less symmetry possibilities owing to the limited twinning operations from respective hosts. Considering the structural similarity of $MoS_2$ and $MoTe_2$ and the STEM images of 1H/1T' domains in $MoS_2$[37], we propose and build three different models for the 1H/1T' boundaries which are named, according to the edge of the 1T' phase, as the Te-terminated (ZZ-Te) and the Mo-terminated (ZZ-Mo) interfaces along the zigzag direction, and the armchair (AC) interface which is aligned along the armchair direction. To compare the stability of each interface, the interface energy of phase boundary ($K_b$) is calculated which a quantity insensitive to the length of the boundary. We show below that the $K_b$ can be derived through varying the size of the 1T' phase. The relaxed structure of 1T' domain with ZZ-Mo boundary is shown in Fig. 2b. $l=nb$ is the boundary length of the triangular 1T' domain where $n$ is the integer number (*n*=3-7) and *b* is the lattice parameter of 1T' along zigzag direction. The formation energy can be expressed as the following formula

$$E_{\text{form}} = E_t - \frac{n(n+1)}{2}E_{T'} - \left(90 - \frac{n(n+1)}{2}\right)E_H + nu_{Te} \quad (1)$$

where $E_t$ is the total energy of the compound system. $E_{T'}$ and $E_H$ are the energies of the 1H and 1T' per chemical unit, respectively. $u_{Te}$ is the chemical potential of tellurium. It is noted that the number of Te atom lost in the ZZ-Mo domain is equal to *n*.

For a triangular domain structure, the formation energy ($E_{\text{form}}$) consists of two parts: the 1H/1T' boundary energy ($E_b$) and the corner energy ($E_c$). For a large-sized 1T' phase, $E_c$ is a constant while $E_b$ is linearly proportional to the size of the boundary ($K_b l$), where $K_b$ is the energy per unit length of the boundary. Thereby, the formation energy is fitted as below,

$$E_{\text{form}} = 3E_b + 3E_c = 3K_b(u_{Te})l + 3E_c(u_{Te}) \quad (2)$$

It is noted that the three boundaries of the triangular domain are different, and therefore $K_b$ could be different for different boundaries. Here, we consider the three edges (corners) as a whole. Since all edges expand or shrink in proportion, their differences are irrelevant and $K_b$ can be considered as the average boundary energy per unit length. For a given chemical potential $u_{Te}$, $K_b$ can be obtained by fitting Eq. (2) to the formation energy vs. size data of the triangle 1T' domain. The calculated formation energies for various sizes of the 1T' domain are shown in Fig. 2c, with $u_{Te}$ taken from bulk Te. From the fitting, we obtained $K_b = 1.50$ eV/nm for the ZZ-Mo boundary. $K_b$ for the ZZ-S and AC boundaries are obtained similarly (see Fig. S7 of Supplementary Information) and the values are 2.37 and 2.82 eV/nm, respectively. Since the formation energy of the ZZ-Mo boundary is much lower than those of the other two boundaries, the ZZ-Mo boundary should prevail for 1T' $MoTe_2$ embedded in 1H $MoTe_2$.

Interestingly, the triangular nanometer-sized fragment of the 1T' domain embedded in the 1H phase behaves as a quantum dot. Due to quantum confinement effect, a metal to semiconductor transition in the 1T' domain takes place as the size of the 1T' fragment is reduced. The HOMO-LUMO gap evaluated at the DFT-GGA level is shown in Fig. 2d and it increases gradually from 31 meV to 240 meV when the domain size is reduced from $l = 2.5$ nm to $l = 1$ nm. A HOMO-LUMO gap in the range of meV corresponds to a Kubo gap δ in the far-infrared region[38]. The monotonic decrease of the gap δ with the domain size is in agreement with the convergence of electronic structure to that of monolayer 1T'

MoTe$_2$, similar to the size-dependent electronic properties found in traditional 3D metallic nanoparticles[39]. Apparently, the gap δ would disappear when the domain size is big enough. The partial charge densities of the HOMO and LOMO are shown in Fig. 2e and 2f, respectively, for $l = 2.1$ nm. The charge distribution is different for the HOMO and the LUMO. The HOMO charge is distributed in the entire 1T' domain while the LUMO charge is primarily located at the vertices of the 1T' domain. Such characteristics of the 2D quantum dots may found applications such as nano-emitters.

**Quantum transmission and states-alignment of homojunction.** To understand the effect of phase boundaries on the electronic transport properties, NEGF simulations were performed to investigate the quantum ballistic tunneling of carriers across the 1H/1T' boundary of MoTe$_2$. We construct asymmetric junctions which include three parts: the semi-infinite left (1T' phase) electrode, the central scattering (1H/1T' interface) region and the right (1H phase) electrode (Fig. 3a). The three types of 1H/1T' boundaries discussed above are investigated. The calculated spatially resolved local densities of states (LDOS) around the interfaces are shown in Fig. 3b. Surprisingly no in-gap defective states are found in the band gap of the 1H phase at the interface which can be ascribed to the overall structural integrity of the 1H/1T' interface. Nevertheless, the semiconductor-metal heterostructure allows a charge spill-over from the metallic phase to the semiconducting phase. This accounts for the band bending in the semiconductor and the Schottky barrier at such a semiconductor-metal interface[40,41]. Indeed, as shown in the LDOS in Fig. 3b, there is a substantial upward band bending (0.75 eV) in the 1H phase in the ZZ-Mo type 1H/1T' MoTe$_2$ interface while less significant bending is found for the other types of interface which indicates that the ZZ-Mo type 1H/1T' boundary favors charge transfer from the 1T' to 1H phase. Our work shows that the band bending, the Schottky barrier height (SBH) and the charge flow in such a 1H/1T' in-plane heterostructure are very sensitive to the interfacial atomic structures.

For the energetically favored lattice registry with 1H/1T' phase boundaries, new atomic bonding and large atomic displacements occur at the boundaries due to the breaking of the periodicity of the host phases. The patterns of atomic displacement at the phase boundaries, relative to their equilibrium positions in respective 1H or 1T' phases, are depicted in Fig. 3c. Large displacements are found in the proximity of the interface, with the strain field in the ZZ-Mo type 1H/1T' boundary attenuating much faster than that in the other two cases. There is no local geometric curvature at the interface which is consistent with the fact that the displacements are largely in plane. The local strain field at the phase boundaries induces a redistribution of charge density and affect the band alignments across the interface.

The transmission spectra of the asymmetric two-probe model are mapped in Fig.

3d at the equilibrium state (zero voltage). One can see that there exists a transmission gap in all cases associated with the semiconducting 1H phase as the electrode. The transmission gap of the ZZ-Te and AC type interfaces is close to 1.1 eV which is equal to the band gap of the 1H phase, while the ZZ-Mo case has a slightly larger transmission gap (about 1.3 eV). The band bending for the ZZ-Mo boundary may account for this larger transmission gap because no states are located between 0.55 and 0.75 eV at the 1H MoTe$_2$ side within 40 Å from the ZZ-Mo interface. In this energy range, electrons from the 1T' phase (left electrode) must overcome a barrier to tunnel into the 1H phase (right electrode), and suffer from more scattering in this process, resulting in a negligible tunneling probability.

**Anisotropic tunneling of homophase junction.** Homophase MoTe$_2$ tunnel junction in the form of 1T'/1H/1T' is highly appealing for single-layered electronic devices. To identify the angular alignment of the junction relative to the host phase, angular dependence of the transport efficiency is examined by aligning the two 1T' leads along both AC- and ZZ- directions (Fig. 4a). For the AC-aligned junction, there are two asymmetric interfaces (ZZ-S and ZZ-Mo interfaces) in the system, while for the ZZ-aligned junction the interfaces are symmetric. To identify the efficiency of the quantum tunneling we have calculated the conductance as a function of the length of the 1H phase in the scattering region. As shown in Fig. 4a, the conductance ($G$) decreases exponentially with the length $L$ of the 1H phase in the central region for both junctions, which can be fitted by the following equation[42]:

$$G(E_f) = G_c e^{[-2\kappa(\Phi,E_f)L]} \qquad (3)$$

where $G_c$ is the contact conductance and $\kappa(\Phi, E_f)$ is the tunneling decay coefficient which is given by $\sqrt{2m_e(\Phi - E_f)}/\hbar$ with $m_e$ the electron mass and $\Phi$ the tunneling barrier. The $G_c$ obtained from the fitting of the AC junction (0.36 $G_0$) is higher than that of the ZZ junction (0.003 $G_0$), due to the larger lattice distortion and more scattering from the interfacial states in the latter. Interestingly, the decay coefficient of the AC junction is 0.25 Å$^{-1}$ which is higher than 0.18 Å$^{-1}$ of the ZZ junction.

As the transport efficiency of the junction is closely related to the intrinsic electronic property of the semiconducting phase in the central region, we investigate the propagating and evanescent states of the 1H phase by calculating its complex band structure. Since the band extrema of 1H MoTe$_2$ are located at the K (1/3, 1/3) point, the tunneling process in principle should be dominated by the evanescent states near the K point because of the lowest tunneling barrier. As seen from the complex band structure (Fig. 4b), evanescent states with the lowest decaying rate $\kappa$ are all located at K point and its symmetrically equivalent replicas, and the value at $E_f$ is 0.19 Å$^{-1}$.

The angular dependence of $\kappa$ at $E_f$ is investigated by constructing a series of rectangular supercells with one lattice direction along the transport direction which is at an angle θ from the ZZ direction of the 1H MoTe$_2$ unit cell (see Fig. S8 Supplementary Information). Location of band maxima in the reciprocal space of each supercell is determined by folding the K point of the unit cell into the Brillouin zone of the supercell (denoted as K'). As shown in Fig. 4c, the lowest $\kappa$ along different directions at K' are the same while the $\kappa$ at Γ appears to have a sixfold symmetry.

In fact, overall transport efficiency relies on a proper matching of evanescent states in the barrier with propagating Bloch states in the electrodes. Herein, transversal $k_x$ states-resolved transmission for the AC and ZZ junctions is depicted in Fig. 4d. The major contribution to the conductance of the ZZ junction arises from $k_x = 0$ which coincides with the K' point (refer to Fig. 4c). In contrast, transmission of the AC junction is dominated by states at the point ($k_x$=0.15, 0) which is slightly away from K' ($k_x$=1/3, 0), despite the smallest $\kappa$ is always located at K' point (Fig. 4c). We also calculated the transversal $k_x$-resolved decaying rate $\kappa$ which is shown in Fig. S9 in Supplementary Information. Interestingly, although we found that the overall $k_x$-transmission profile is consistent with the trend of the $k_x$-$\kappa$ relationship in Fig. S9 of Supplementary Information, that is, a smaller $\kappa$ leads to a higher transmission, the location of the transmission peak in the AC junction slightly deviates from the valley position of $k_x$-$\kappa$ curve. This can be traced back to the fact that the conductance is not solely dominated by the damping rate in the central 1H phase but also related to the interfacial matching of the propagating Bloch states in the 1T' phase and the evanescent states in the 1H phase. Overall, our results demonstrate that such homophase junctions could exhibit anisotropic transport characteristics owing to the anisotropic $\kappa$ and interfacial matching albeit host MoTe$_2$ has an isotropic structure.

**Discussion**

The interplay between charge, spin and orbital results in extremely rich crystal phases (i.e. 1H, 1T and 1T') and quantum excitions such as charge density waves (CDW) and spin density waves (SDW) in TMDs materials, which enable versatile applications in optoelectronics and valleytronics[43-45]. Existence of ubiquitous inhomogeneous electron-hole puddles within a single sheet of the TMDs triggers the coexistence of multiple phases in a single sheet, and thereafter the formation of various interfaces allowing molecular patterning[46]. While the transition can be triggered by application of a strain[30], our work shows that the atomically thin lattice is highly sensitive to the density of excess carriers, i. e., the 1T'-1H phase transition temperature of MoTe$_2$ decreases linearly with both electron and hole doping, and drops to 300 K with a small hole doping about 0.08 $q_0$ and electron doping about 0.04 $q_0$, equivalent to 0.08 hole and 0.04 electron doped per unit cell, respectively. Note our predicted transition temperature of undoped MoTe$_2$ is 1200 K as shown in Fig. S2 in Supplementary Information, slightly higher than the experimental value of 1100K[30].

We predict that quantum confinement and injection of carriers in 1T' MoTe$_2$ leads to the formation of Kubo gap which was only observed in 3D metallic nanoclusters such as nano gold clusters previously. The finite Kubo gap in nanosized 1T' phase of MoTe$_2$, a metallic droplet squeezed in atomically thin 2D sheet, implies a strong quantum confinement effect. Previous studies showed that this ultimate confinement leads to a strong exciton binding energy in TMDs[47]. Here the opening of a finite gap in the nanosized metallic phase, i.e. 1T' MoTe$_2$, may lead to a size-dependent plasmonic oscillation of carriers which can serve as efficient quantum emitters. Patterned 1T'/1H phase with metallic-semiconducting periodicity may create metamaterials with novel plasmon-polariton interactions and allow unique spin manipulation owing to its largely coherent interfaces.

**Methods**

**Electronic structure calculations.** The ground state electronic structure calculations are performed by using the Vienna ab initio Simulation Package (VASP)[48] within the framework of DFT. We use projected augmented wave together with cutoff energy of 400 eV. The generalized gradient approximation (GGA) in the Perdew–Burke–Ernzerhof form is chosen as exchange correlation functional. In addition, the Brillouin zone is sampled by an 11×21×1 $k$-point grid. The convergence criterion of total energy is set to be $10^{-6}$ eV. All the structures are fully relaxed until the forces on atoms are smaller than 0.01 eV/Å. The kinetics analysis of the activation barrier of the 1H/1T' structural phase transformation is conducted by the CI-NEB calculation.

**Phononic and thermodynamics calculation.** To evaluate the stability of structures under different doped and strained conditions, phonon dispersion is calculated by using the displacement method. The second order force constants are obtained with calculating the energies of displaced configurations based on a 3×3×1 supercell together with a 6×6×1 $k$-point grid. Based on the obtained Hessian matrices, quasi-harmonic vibrational free-energy correction is added to the total energy evaluated by the DFT and the Helmholtz free energy is calculated as

$$F(T) = \sum_{qj} \frac{1}{2} \hbar \omega_{qj} + k_B T \sum_{qj} \ln[1 - \exp(-\hbar\omega_{qj}/k_B T)] \quad (4)$$

where $\omega_{qj}$, $\hbar$, $T$ and $k_B$ are the frequencies, the reduced Planck constant, temperature and Boltzmann constant, respectively. $q$ and $j$ denote the momentum and mode index of the phonon.

**Quantum ballistic transport analysis by NEGF.** The transport calculations in the ballistic region are performed by using Atomistix Tool Kit which is based on DFT combined with NEGF method[49]. The OPENMX norm-conserving pseudopotentials with GGA are adopted with 100 Hartree density mesh cutoff for the grid integration. Pseudoatomic orbitals basis sets $s^3p^2d^1$ and $s^2p^2d^1$ are utilized for Mo and Te atoms, respectively. Double-ζ-polarized basis is adopted in transport calculations. In the metallic tunneling junction with the 1T'-1H-1T' configuration, the energy ($E$) and the transversal momentum ($k_{//}$) resolved ballistic transmission ($T$) at equilibrium of spin σ component is calculated by the formula:

$$T_\sigma(E, k_{//}) = \mathrm{Tr}[\Gamma_L(E, k_{//}) G_\sigma^r(E, k_{//}) \Gamma_R(E, k_{//}) G_\sigma^a(E, k_{//})] \quad (5)$$

where $\Gamma_{L/R}$ is the self-energy item which takes the coupling of the central scattering region and the left (L)/right (R) electrode into consideration, and $G_\sigma^{r/a}(E, k_{//})$ is the retarded/advanced Green's function matrix of the system. Along the transporting direction, the zero-bias quantum conductance $G$ is computed as

$$G = \frac{e^2}{h} \sum_{k_{//}, \sigma} T_\sigma(E, k_{//}) \quad (6)$$

In this work, the $G$ along AC and ZZ directions of MoTe$_2$ is evaluated by integrating the electron transmission at the $E_\text{f}$ for over $k_{//}$ states along the transverse direction, respectively.

**Additional information**

Supplementary Information is available online.

Competing interests: The authors declare no competing financial interests.

**References**

1. Jena, P. & Castleman, A.W., Clusters: A bridge across the disciplines of physics and chemistry. *Proc. Natl. Acad. Sci. U S A* **103**, 10560-10569 (2006).

2. Kubo, R. Electronic properties of metallic fine particles. *I. J. Phys. Soc. Jpn.* **17**, 975-986 (1962).

3. Issendorff, B. V. & Cheshnovsky, O. Metal to insulator transitions in clusters. *Annu. Rev. Phys. Chem.* **56**, 549 (2005).

4. Ramachandra Rao, C. N., Kulkarni, Giridhar U., John Thomas, P. & Edwards, Peter P. Metal nanoparticles and their assemblies. *Chem. Soc. Rev.* **29**, 27–35 (2000).

5. Higaki, T., Zhou, M., Lambright, K. J., Kirschbaum, K., Sfeir, M. Y. & Jin, R. Sharp transition from nonmetallic Au$_{246}$ to metallic Au$_{279}$ with nascent surface plasmon resonance. *J. Am. Chem. Soc.* **140**, 5691-5695 (2018).

6. Baletto, F. & Ferrando, R. Structural properties of nanoclusters: Energetic, thermodynamic, and kinetic effects. *Rev. Mod. Phys.* **77**, 371 (2005).

7. Moseler, M., Häkkinen, H., Barnett, R. N. & Landman, U. Structure and magnetism of neutral and anionic palladium clusters. *Phy. Rev. Lett.* **86**, 2545 (2001).

8. Nanda, K. K. On the paradoxical relation between the melting temperature and forbidden energy gap of nanoparticles. *J. Chem. Phys.* **133**, 054502 (2010)

9. Liu, C., Li, T., Abroshan, H., Li, Z., Zhang, C., Kim, H.J., Li, G. & Jin, R. Chiral $Ag_{23}$ nanocluster with open shell electronic structure and helical face-centered cubic framework. *Nat. Commun.* **9**, 744 (2018).

10. Weng, B., Lu, K. Q., Tang, Z., Chen, H. M. & Xu, Y. J. Stabilizing ultrasmall Au clusters for enhanced photoredox catalysis. *Nat. Commun.* **9**, 1543 (2018).

11. Gao, M., Lyalin, A., Takagi, M., Maeda, S. & Taketsugu, T. Reactivity of gold clusters in the regime of structural fluxionality. *J. Phys. Chem. C* **119**, 11120-11130 (2015).

12. Billas, I. M., Chatelain, A. & de Heer, W. A. Magnetism from the atom to the bulk in iron, cobalt, and nickel clusters. *Science* **265**, 1682-1684 (1994).

13. Yoo, Y., DeGregorio, Z. P., Su, Y., et al. In-Plane 1H-1T' $MoTe_2$ homojunctions synthesized by flux-controlled phase engineering. *Adv. Mater.* **29**, 1605461 (2017).

14. Wang, Y., Xiao, J., Zhu, H., Li, Y., Alsaid, Y., Fong, K. Y., Zhou, Y., Wang, S., Shi, W., Wang, Y. & Zettl, A. Structural phase transition in monolayer $MoTe_2$ driven by electrostatic doping. *Nature* **550**, 487 (2017).

15. Zhang, X., Jin, Z., Wang, L., Hachtel, J. A., Villarreal, E., Wang, Z., Ha, T., Nakanishi, Y., Tiwary, C. S., Lai, J. & Dong, L. Low contact barrier in 1H/1T' $MoTe_2$ in-plane heterostructure synthesized by chemical vapor deposition. *ACS Appl. Mater. Inter.* **11**, 12777-12785 (2019).

16. Yang, H., Kim, S. W., Chhowalla, M. & Lee, Y. H. Structural and quantum-state phase transitions in van der Waals layered materials. *Nat. Phys.* **13**, 931 (2017).

17. Radisavljevic, B., Radenovic, A., Brivio, J., et al. Single-layer $MoS_2$ transistors. *Nat. Nanotech.* **6**, 147-150 (2011).

18. Wang, W., Narayan, A., Tang, L., et al. Spin-valve effect in $NiFe/MoS_2/NiFe$ junctions. *Nano Lett.* **15**, 5261-5267 (2015).

19. Zeng, H., Dai, J., Yao, W., et al. Valley polarization in $MoS_2$ monolayers by optical pumping. *Nat. Nanotech.* **7**, 490-493 (2012).

20. Wang, Q. H., Kalantar-Zadeh K., Kis, A., et al. Electronics and optoelectronics of two-dimensional transition metal dichalcogenides. *Nat. Nanotech.* **7**, 699-712 (2012).

21. Ali, M. N., Xiong, J., Flynn, S., et al. Large, non-saturating magnetoresistance in $WTe_2$. *Nature* **514**, 205-208 (2014).

22. Qian, X., Liu, J., Fu, L., et al. Quantum spin Hall effect in two-dimensional transition metal dichalcogenides. *Science* **346**, 1344-1347 (2014).

23. Lukowski, M. A., Daniel, A. S., Meng, F., et al. Enhanced hydrogen evolution catalysis from chemically exfoliated metallic $MoS_2$ nanosheets. *J. Am. Chem. Soc.* 135, 10274-10277 (2013).

24. Gao, G., Jiao, Y., Ma, F., et al. Charge mediated semiconducting-to-metallic phase transition in molybdenum disulfide monolayer and hydrogen evolution reaction in new 1T′ phase. *J. Phys. Chem. C* **119**, 13124-13128 (2015).

25. Keum, D. H., Cho, S., Kim, J. H., et al. Bandgap opening in few-layered monoclinic $MoTe_2$. *Nat. Phys.* **11**, 482-486 (2015).

26. Duerloo, K. A. N., Li, Y. & Reed, E. J. Structural phase transitions in two-dimensional Mo-and W-dichalcogenide monolayers. *Nat. Commun.* **5**, 5214 (2014).

27. Cho, S., Kim, S., Kim, J. H., et al. Phase patterning for ohmic homojunction contact in MoTe$_2$. *Science* **349**, 625-628 (2015).

28. Si, C., Choe, D., Xie, W., Wang, H., Sun, Z., Bang, J. & Zhang, S., Photoinduced vacancy ordering and phase transition in MoTe$_2$. *Nano Lett.* **196**, 3612-3617 (2019).

29. Kim S, Song S, Park J, et al. Long-range lattice engineering of MoTe$_2$ by 2D electride. *Nano Lett.* **17**, 3363-3368 (2017).

30. Song S., Keum, D. H., Cho, S., et al. Room temperature semiconductor–metal transition of MoTe$_2$ thin films engineered by strain. *Nano Lett*. **16**, 188-193 (2015).

31. Rhodes, D., Chenet, D. A., Janicek, B. E., et al. Engineering the structural and electronic phases of MoTe$_2$ through W substitution. *Nano Lett.* **17**, 1616-1622 (2017).

32. Mounet, N., Marzari, N. First-principles determination of the structural, vibrational and thermodynamic properties of diamond, graphite, and derivatives. *Phys. Rev. B* **71**, 205214 (2005).

33. Wang, Y., Xiao, J., Zhu, H., et al. Structural phase transition in monolayer MoTe$_2$ driven by electrostatic doping. *Nature* **550**, 487 (2017).

34. Li, W. & Li, J. Ferroelasticity and domain physics in two-dimensional transition metal dichalcogenide monolayers. *Nat. Communi.* **7**, 10843 (2016).

35. Cooper, R. C., Lee, C., Marianetti, C. A., et al. Nonlinear elastic behavior of two-dimensional molybdenum disulfide. *Phys. Rev. B* **87**, 035423 (2013).

36. Eda, G., Fujita, T., Yamaguchi, H., et al. Coherent atomic and electronic heterostructures of single-layer MoS$_2$. *ACS Nano* **6**, 7311-7317 (2012).

37. Lin, Y. C., Dumcenco, D. O., Huang, Y. S., et al. Atomic mechanism of the semiconducting-to-metallic phase transition in single-layered MoS$_2$. *Nat. Nanotech.* **9**, 391-396 (2014).

38. Ghosh, S. K. Kubo gap as a factor governing the emergence of new physicochemical characteristics of the small metallic particulates. *Assam Univ. J. Sci. Technol.* **7**, 114-121 (2011).

39. Häberlen, O. D., Chung, S. C., Stener, M., et al. From clusters to bulk: A relativistic density functional investigation on a series of gold clusters $Au_n$, n= 6,…, 147. *J. Chem. Phys.* **106**, 5189-5201 (1997).

40. Bardeen, J. Surface states and rectification at a metal semi-conductor contact. *Phys. Rev.* **71**, 717 (1947).

41. Mönch, W. Role of virtual gap states and defects in metal-semiconductor contacts. *Phys. Rev. Lett.* **58**, 1260 (1987).

42. Mavropoulos, P., Papanikolaou, N., Dederichs, P. H. Complex band structure and tunneling through ferromagnet/insulator/ferromagnet junctions. *Phys. Rev. Lett.* **85**, 1088 (2000).

43. Joe, Y. I., Chen, X. M., Ghaemi, P., et al. Emergence of charge density wave domain walls above the superconducting dome in $1T\text{-}TiSe_2$. *Nat. Phys.* **10**, 421-425 (2014).

44. Katagiri, Y., Nakamura, T., Ishii, A., et al. Gate-tunable atomically thin lateral $MoS_2$ Schottky junction patterned by electron beam. *Nano Lett.* **16**, 3788-3794 (2016).

45. Yin, X., Wang, Q., Cao, L., et al. Tunable inverted gap in monolayer quasi-metallic $MoS_2$ induced by strong charge-lattice coupling. *Nat. Commun.* **8**, 486 (2017).

46. Lin, X., Lu, J. C., Shao, Y., et al. Intrinsically patterned two-dimensional materials for selective adsorption of molecules and nanoclusters. *Nat. Mater.* **16**, 717 (2017).

47. Eda, G., Yamaguchi, H., Voiry, D., et al. Photoluminescence from chemically exfoliated $MoS_2$. *Nano Lett.* **11**, 5111-5116 (2011).

48. Kresse, G., Furthmüller, J. Efficiency of ab-initio total energy calculations for metals and semiconductors using a plane-wave basis set. *Comput. Mater. Sci.*, **6**, 15-50 (1996).

49. Soler, J. M., Artacho, E., Gale, J. D., et al. The SIESTA method for ab initio order-N materials simulation. *J. Condens. Matter Phys.* **14**, 2745 (2002).

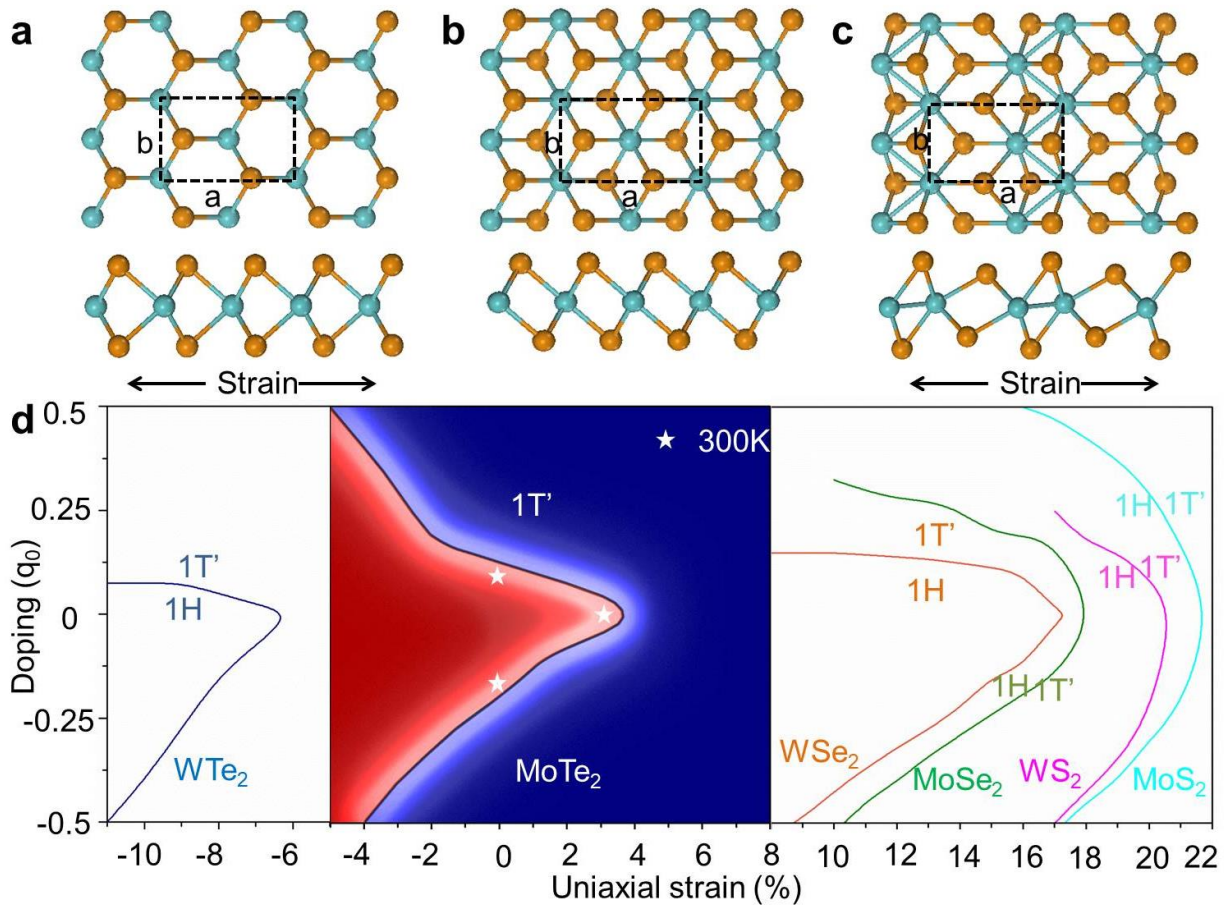

**Figure 1. Phase diagram of isolated MX$_2$ (M: Mo or W and X: S, Se, or Te) under strain and excess charges.** The polytypic structures of MX$_2$: (a) 1H, (b) 1T and (c) 1T'. Blue spheres are M or W atoms, and orange spheres are X atoms. (d) The phase diagrams of six VI group TMDS as a function of both charge doping and uniaxial strain along the long rectangular lattice. The star points represent the phase transition at room temperature with the vibrational free energy correction under quasi-harmonic approximation.

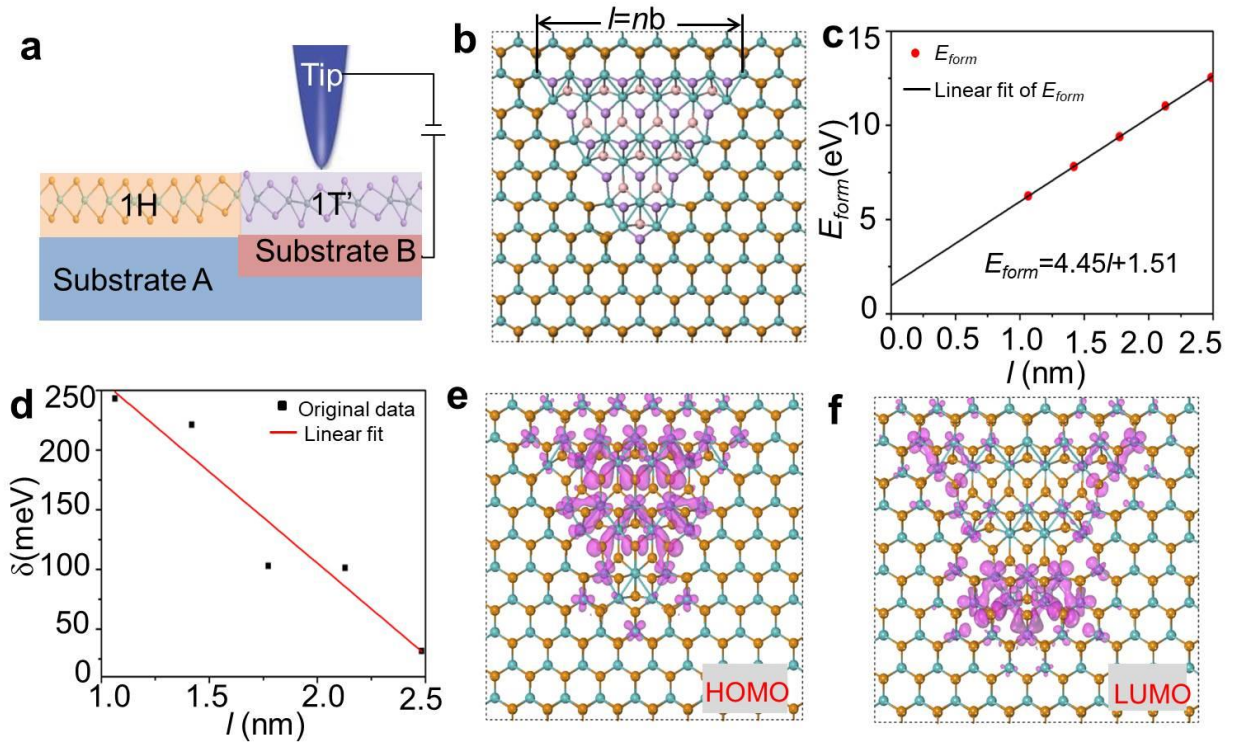

**Figure 2. Formation and scaling behavior of Kubo gap δ in 1H/1T' MoTe$_2$ nanodomain.** The schematic of local phase transition induced by charge doping and strain (a), structure of triangular 1T' domain in 1H phase with ZZ-Mo boundary (b), the formation energies of triangular 1T' domain with different sizes(c), The size dependent δ gap(d). The charge distribution of HOMO and LUMO of a 1T' domain (e and f).

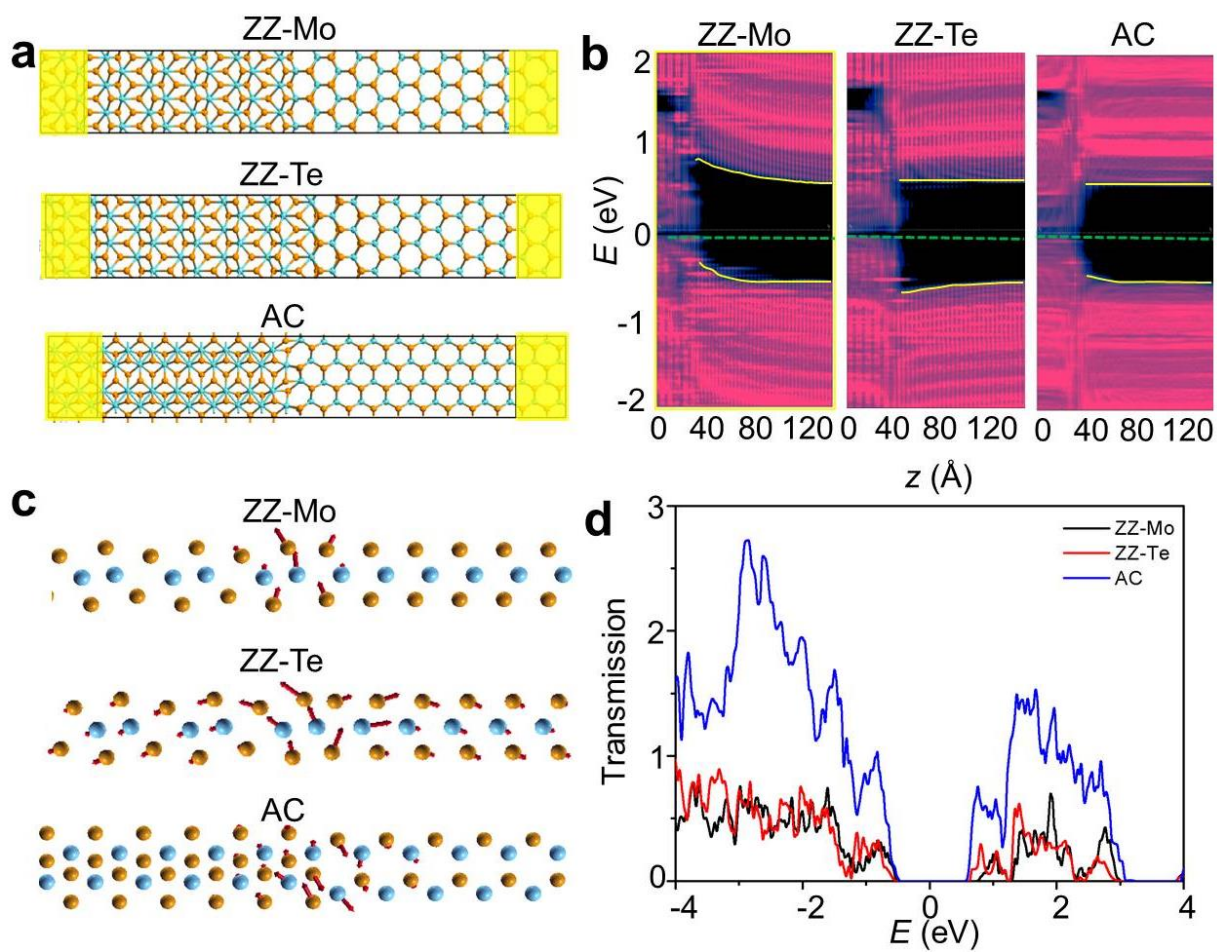

**Figure 3. Atomic adjustments and electronic renormalization at interfaces in 1H/1T' MoTe$_2$ homojunction.** (a) Schematic atomic model for the different phase boundaries of 1H/1T': ZZ-Mo, ZZ-Te, AC. (b) Spatially resolved DOS in the proximity of the boundaries. The energy is relative to the Fermi level ($E_f$) and the color bar denotes the amplitude of DOS. (c) The length and head of the arrows denote the magnitude and direction of the displacement of the corresponding atoms. (d) Transmission spectra of the asymmetric junction with 1H (1T') phase as the right (left) electrode.

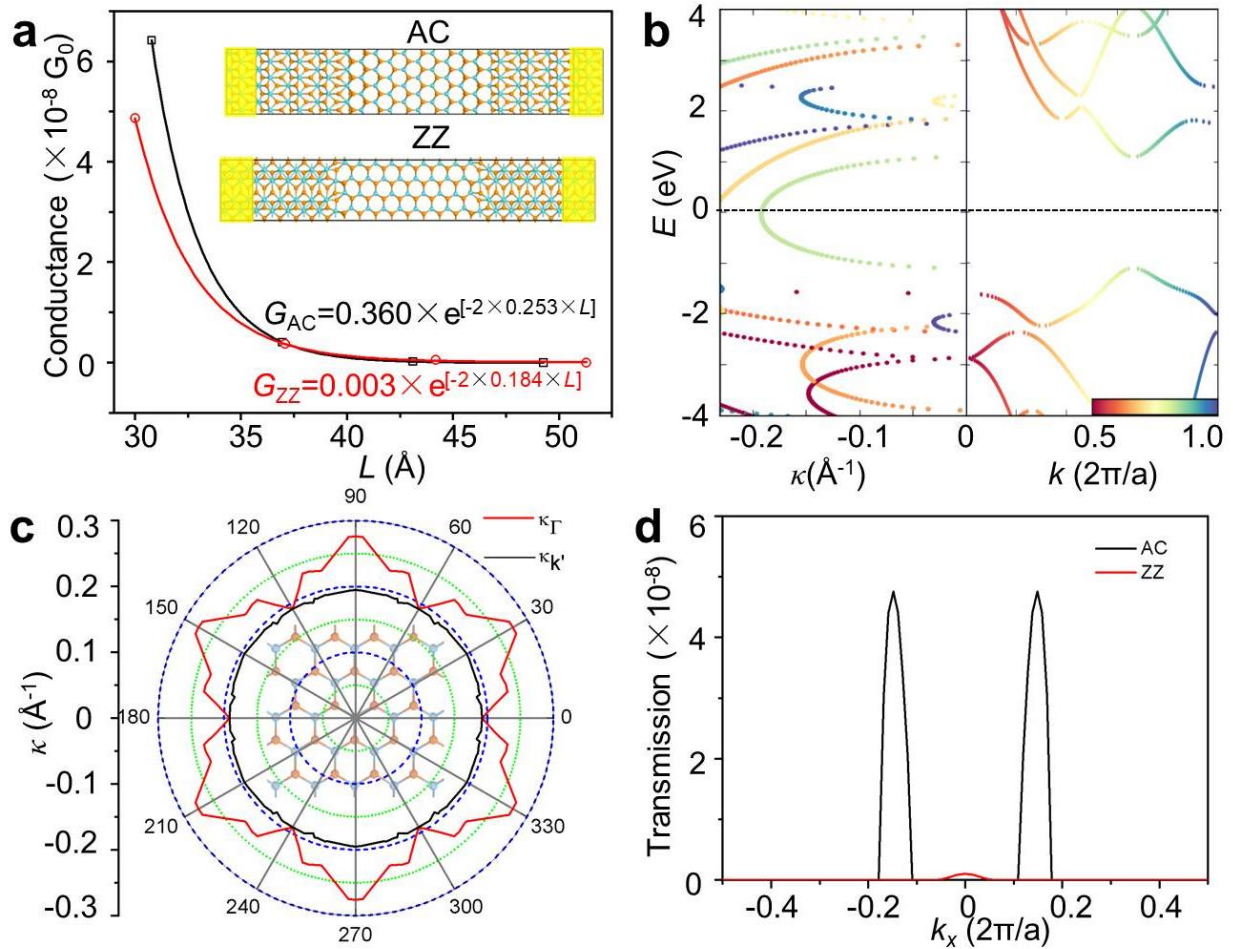

**Figure 4. Orientation effect of electronic injection and damping across 1H/1T' MoTe₂ homojunction.** (a) Conductance $G$ (in units of $G_0=2e^2/h$) of 1T'/1H/1T' junctions at $E_f$ as a function of length $L$ of the central 1H MoTe₂ phase along AC and ZZ directions. (b) Complex band structure of 1H MoTe₂ along ZZ direction at the $\Gamma$ point. he real (right panel) and imaginary (left panel) parts represent propagating and evanescent states, respectively. (c) Polar plot of the lowest value of $\kappa$ at $E_f$ for both $\Gamma$ and K' along different directions. Note zero degree corresponds to ZZ direction. (d) Transversal $k_x$ resolved transmission of the junctions along AC and ZZ direction.